\documentclass[review]{elsarticle}

\usepackage{graphics}
\usepackage{graphicx}
\usepackage{epsfig}
\usepackage{amssymb}

\journal{Elsevier}
\begin{document}
\begin{frontmatter}


\title{Dynamic phase transitions in a ferromagnetic thin film
system: A Monte Carlo simulation study}
\author[1,2]{Erol Vatansever\corref{cor1}}
\cortext[cor1]{Corresponding author. Tel.: +90 3019547; fax:
+90 2324534188.} \ead{erol.vatansever@deu.edu.tr}
\author[1]{Hamza Polat}
\address[1]{Department of Physics, Dokuz Eyl\"{u}l University,
Tr-35160 \.{I}zmir, Turkey}
\address[2]{Dokuz Eyl\"{u}l University, Graduate School of Natural
and Applied Sciences, Turkey}


\begin{abstract}
Dynamic phase transition properties of ferromagnetic thin film system under the influence
both bias and time dependent magnetic fields have been elucidated by means of kinetic
Monte Carlo simulation with local spin update Metropolis algorithm. The obtained results
after a detailed analysis suggest that bias field is the conjugate field to  dynamic
order parameter,  and  it also appears to define a phase line between two antiparallel
dynamic ordered states depending on the considered system parameters.
Moreover, the data presented in this study  well qualitatively reproduce
the recently published experimental findings  where time dependent magnetic
behavior of a uniaxial cobalt films is  studied in the neighborhood of
dynamic phase transition point.
\end{abstract}

\end{frontmatter}
\section{Introduction}\label{introduction}
\hspace{0.4cm}  When a magnetically interacting spin system with ferromagnetic coupling $J$ is
exposed to a magnetic field oscillating in time, the system may not respond to the
external perturbation simultaneously, and hereby, two important striking phenomena occur:
Non-equilibrium phase transitions and dynamic hysteresis behavior. A typical ferromagnetic
system exists in dynamically disordered phase, in which the time dependent magnetization of the system
oscillates around zero value in the range of high temperature and high amplitude of field.
In contrary to the aforementioned treatment, for the low temperature and small amplitude of
field regimes, the time dependent magnetization exhibits an oscillating behavior in a restricted
range around one of its two non-zero values. For the first time, this type of investigation
regarding the dynamic phase transition properties of Ising model under the presence of
a sinusoidally oscillating magnetic field have been performed by  Tom\'{e} and Oliveira
by benefiting from Molecular Field Theory (MFT) \cite{Tome}. Since then, a great deal works
concerning the non-equilibrium phase transitions as well as hysteresis behaviors of different
types of magnetic systems have been investigated by a variety of techniques
such as MFT \cite{Acharyya1, Buendia, Keskin1, Canko, Punya, Ertas1}, Effective Field Theory
(EFT) \cite{Shi, Deviren, Ertas2, Aktas, Yuksel1, Vatansever1, Akinci},
and Monte Carlo simulations (MC) \cite{Acharyya1, Acharyya2, Acharyya3, Sides,
Laosiritaworn, Park, Yuksel2, Vatansever2, Vatansever3}.
For example, thermal and magnetic phase transition properties of the
kinetic Ising model  have been analyzed within the framework of MFT
in \cite{Punya}, and it is found that frequency dispersion of the hysteresis
loop area, the remanence and the coercivity have been categorized into
three distinct types for varying system parameters. Furthermore,  Park and Pleimling have
considered kinetic Ising models with surfaces subjected to a periodic
oscillating magnetic field to probe the  role of surfaces at dynamic phase transitions.
They have reported that the non-equilibrium surface universality class differs from that of the
equilibrium system, although the same universality class prevails
for the corresponding bulk systems \cite{Park}.

It is obvious from the above picture that only influences of a time
dependent magnetic field  on the dynamic phase transitions properties of the
kinetic Ising model and its derivations have been realized, and possible
situations have been addressed by means of theory and modeling.
Besides, as far as we know, there are a rather few experimental and theoretical
studies associated with the discussion of possible role of an additionally time-independent magnetic
field, namely bias field $h_{b}$, on the  dynamic phase transitions in the
magnetic systems \cite{Robb1, Robb2, Idigoras, Gallardo, Yuksel3}, and these studies
suggest that $h_{b}$ appears to be conjugate field of the dynamic
order parameter. Very recently, this fact has been verified by experimentally
and theoretically in Ref. \cite{Berger} where time dependent magnetic behavior of a uniaxial cobalt films under
the presence of both bias and time dependent magnetic fields
has been studied, and after a detailed analysis, it is observed that the bias field  is
conjugate field of  the dynamic order parameter. Berger and co-workers in Ref. \cite{Berger}
have used the MFT as a theoretical tool. It is a fact that spin fluctuations are ignored
and the obtained results do not have any  microscopic information details of system in MFT.
Keeping in this mind,  in order to qualitatively reproduce the experimental observations,
we have implemented a series of MC simulations of a
ferromagnetic thin film system described by a simple Ising
Hamiltonian including both bias and time dependent oscillating
magnetic fields. Some outstanding results  are given in this letter,
and it can be said that our findings qualitatively support and confirm the experimental
results \cite{Berger}.

The paper organized as follows: In section \ref{formulation},
we briefly introduce our model and MC simulation procedure.
The results and discussion are presented in
section \ref{discussion},  and finally section \ref{conclusion}
includes our conclusions.

\section{Formulation}\label{formulation}
\hspace{0.4cm} We consider a ferromagnetic thin film with
thickness $L_{z}$ under the existence of both bias and time dependent magnetic fields.
The Hamiltonian of the considered system  can be written in the following form:

\begin{equation}\label{eq1}
H=-\sum_{\langle ij \rangle}J_{ij} S_{i}S_{j}-h(t)\sum_{i}S_{i},
\end{equation}
where $S_{i}$ is conventional Ising spin variable which can take values of $S_{i}=\pm1$, and $J_{ij}$ is the
nearest-neighbor spin-spin interaction term, and it is kept fixed as $J_{ij}=J(>0)$ in throughout the
simulation. The first summation in Eq. (\ref{eq1}) is over the nearest-neighbor site pairs
while the second one is over all lattice sites in the thin film system. $h(t)$ denotes the time dependent
oscillating magnetic field term, and it composes of both bias and time dependent oscillating magnetic fields
which has the following form:
\begin{equation}
h(t)=h_{b}+h_{0}\sin(\omega t),
\end{equation}
here $h_{b}$ is the bias field, $t$ is time, $h_{0}$ and $\omega$
are amplitude and angular frequency of the driving magnetic field,
respectively. The period of the oscillating magnetic field is
given by $\tau=2\pi/\omega$.

We simulate the system specified by the Hamiltonian in Eq. (\ref{eq1}) on
a $L_{x} \times L_{y} \times L_{z}$ simple cubic lattice, and we apply free boundary
condition in the $z-$ direction of the thin film system, whereas in the directions perpendicular
to the $z-$ direction  we use periodic boundary conditions. We have studied the thin
film system with thickness $L_{z}=20$ with $L_{x}=L_{y}=100$. The simulation
procedure we follow in this work is as following: The simulation begins at a
high temperature using a random initial condition, and then the system is slowly cooled down with a
reduced temperature step $k_{B}\Delta T/J=0.03$, where $k_{B}$ and $T$ are the
Boltzmann constant and temperature, respectively. The configurations were
generated by selecting the sites sequentially  through the lattice and making single-spin-flip
attempts,  which were accepted or rejected according to the Metropolis
algorithm \cite{Binder, Newman}.  The numerical data were generated over 50 independent sample
realizations by running  the simulations for 20000 MC steps per site
after discarding the first 10000 MC steps. This amount of transient steps
is found to be sufficient for thermalization for the whole range of the
parameter sets.

Our program calculates the instantaneous value of the total magnetization $M(t)$ at
time t as following:
\begin{equation}\label{eq2}
 M(t)=\frac{N_{S}M_{S}(t)+N_{B}M_{B}(t)}{N_{S}+N_{B}},
\end{equation}
where $M_{S}(t)=\frac{1}{N_{S}}\sum_{i=1}^{N_{S}} S{i}$ and  $M_{B}(t)=\frac{1}{N_{B}}\sum_{i=1}^{N_{B}} S_{i}$ while
$N_{S}(=\L_{x}\times L_{y}\times 2)$ and $N_{B}(=\L_{x}\times L_{y}\times (L_{z}-2))$ terms  correspond to the total
number of the spins located on the  surface and  bulk of the thin film system, respectively.
In this study, we are interested only in thermal variation of  total dynamic order
parameter as a function of system parameters, hence,
from the instantaneous magnetization, we can obtain the dynamic order parameter as follows \cite{Tome}:
\begin{equation}\label{eq3}
Q=\frac{1}{\tau}\oint M(t)dt.
\end{equation}
In order to specify the dynamic phase transition point at which dynamically ferromagnetic and paramagnetic phases
separate from each other, we use and check the thermal variation of dynamic
heat capacity which is defined according to the following equation:
\begin{equation}\label{eq4}
C=\frac{dE}{dT},
\end{equation}
where $E$ is the energy per spin over a full cycle of the external applied magnetic field which has the following form:
\begin{equation}\label{eq5}
E=-\frac{1}{\tau(N_{S}+N_{B})}\oint H dt.
\end{equation}
\section{Results and Discussion}\label{discussion}

\hspace{0.4cm} In this section, first of all, in order to understand the dynamic evolution
of the magnetic system in detail, we will focus our attention on non-equilibrium
phase transition properties of the ferromagnetic thin film system
under the existence of only a time dependent oscillating magnetic field.
We will argue and discuss how the amplitude and period of the field affect
the dynamic critical nature of the system. Next,  for some selected combinations
of Hamiltonian parameters, we will give and examine the bias field effects on the
thermal and magnetic properties of the thin film system. Finally, we will discuss
the competing mechanism between bias and time dependent magnetic fields.

\begin{figure}[!here]
\begin{center}
\includegraphics[width=9.5cm,height=7.5cm]{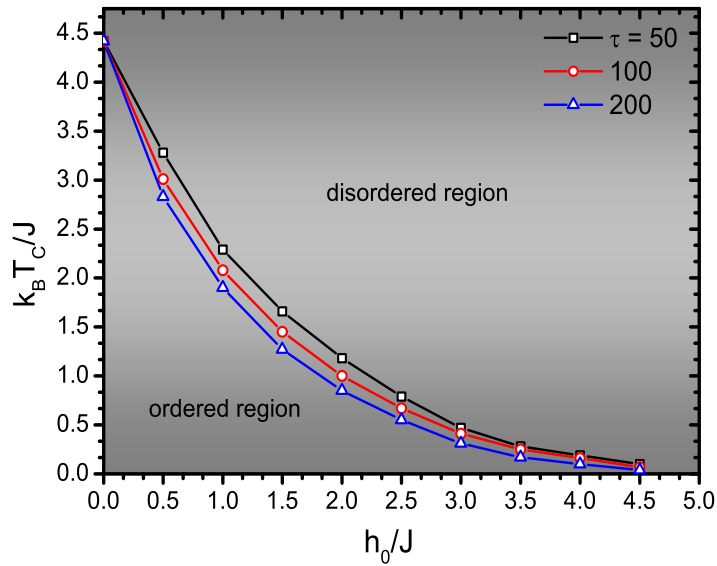}
\caption{Dynamic phase diagram in a $(h_{0}/J-k_{B}T_{C}/J)$ plane obtained from the peaks of
the dynamic heat capacity for selected values of applied field period
such as $\tau=50, 100$ and $200$.}\label{Fig1}
\end{center}
\end{figure}

Dynamic phase diagram separating dynamically ordered and disordered phases
of the thin film system  in a $(h_{0}/J-k_{B}T_{C}/J)$ plane
in the absence  of bias field is shown in Fig. \ref{Fig1} at various $\tau$ values.
At first sight, it can be easily seen from the figure that, for a fixed value of $\tau$,
the system shows ordered phase at the relatively low temperature and applied field amplitude
regions. Here, the time dependent magnetization of the system oscillate around a
non-zero value, namely, the system may not respond to the external magnetic field instantaneously.
With increasing strength of the amplitude of the external field,
the magnetic phase of the system tends to shift dynamically paramagnetic phase.
In this region, the time dependent magnetization of the thin film system oscillate around a
zero value, and it can follow the external field with a relatively
small phase lag depending on the studied system parameters.
One of the obtained results of our MC simulation study is that the thin film system shows
sensitivity to varying applied field period in accordance with the expectations.
For a fixed value of $h_{0}/J$, it is obvious that as $\tau$ is increased, the phase transition point is
lowered because of the fact that decreasing field frequency leads to a decreasing
phase delay between the magnetization and magnetic field  and this makes the
occurrence of the dynamic phase transition  easy. The physical discussions
mentioned above can be easily visualized  by checking any two values of
oscillation periods such as $\tau=50$ and $200$ for fixed value of $h_0/J=1.5$.

\begin{figure}[!here]
\begin{center}
\includegraphics[width=9.5cm,height=7.5cm]{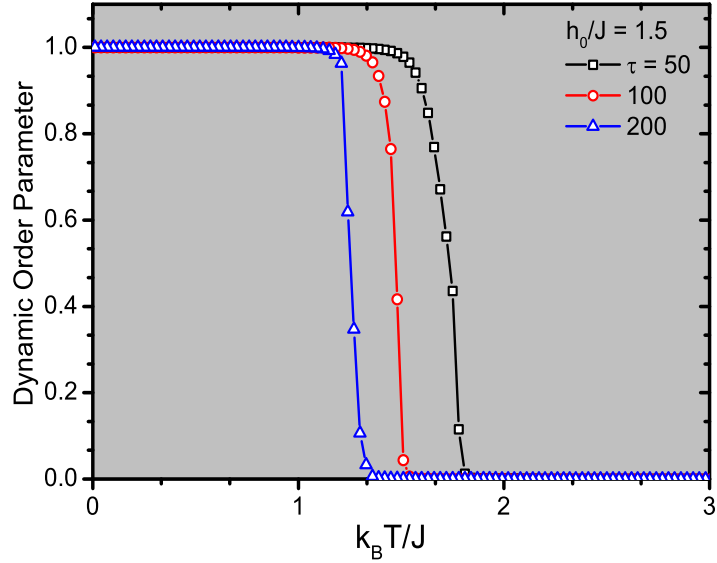}
\caption{Effects of the external applied field period on the temperature versus dynamic order parameter
curves corresponding the dynamic phase diagram illustrated in Fig. \ref{Fig1}.}\label{Fig2}
\end{center}
\end{figure}

In Fig. \ref{Fig2} we present the thermal variations of dynamic order parameter of thin film
system corresponding to phase diagrams plotted in Fig. \ref{Fig1}  for various $\tau$ values
with a selected applied field amplitude $h_{0}/J=1.5$. It is evident from
our simulation that when the temperature increases starting from zero, the
dynamic order parameter starts to decrease from its saturation
value, and it undergoes a second order dynamic phase transition between dynamically ordered and
disordered phases. As we discussed above, dynamic phase transition point moves downward in
the temperature space with increasing applied field period. It is possible to see more clearly these facts
in this figure.

\begin{figure}[!here]
\begin{center}
\includegraphics[width=10.0cm,height=7.5cm]{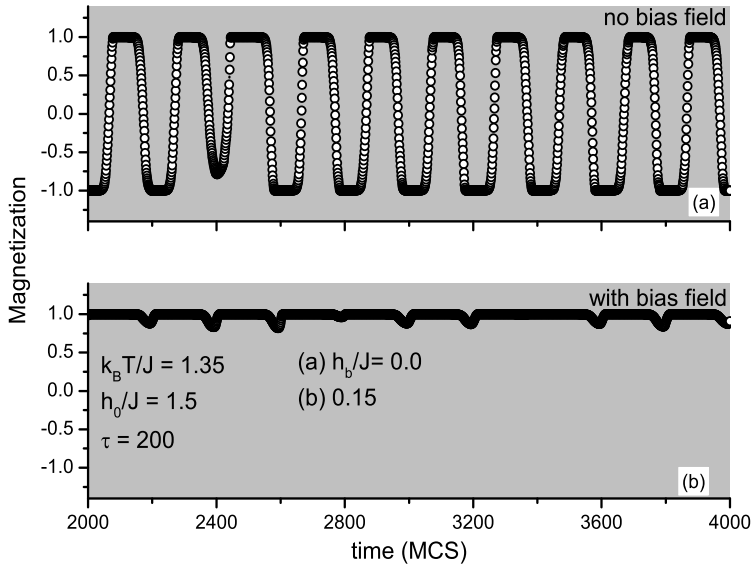}
\caption{For considered  values of $h_{0}/J=1.5, k_{B}T/J=1.35$ and $\tau=200$,
time series of magnetization of ferromagnetic thin film system.  (a) and (b) are given for $h_{b}/J=0.0$
and $0.15$, respectively.
}\label{Fig3}
\end{center}
\end{figure}

Before we start to discuss the influences of bias field on the
thermal and magnetic behavior of the thin film system, it is beneficial
to underline that there exists a competition between bias and time
dependent magnetic fields. Namely, the bias field tries to
keep the system in ordered phase $(Q\neq0) $ whereas the time dependent
magnetic field tries to drive the system into dynamically
disordered phase $(Q=0) $ depending on the other system parameters.
From this point of view,  in Fig. \ref{Fig3} (a) we give
time series of the  magnetization without a bias field for selected  values of
$h_{0}/J=1.5, k_{B}T/J=1.35$ and $\tau=200$. One can deduce
from the figure that the unbiased magnetization curve displays a nearly
rectangular shaped time sequence around zero value such that
the thin film system  exists in dynamically
paramagnetic phase. As shown in Fig. \ref{Fig3}(b), by keeping the system parameters fixed,
if one applies the bias field to the thin film system, for example $h_{b}/J=0.15$,
the biased magnetization curve shows a small variation around its saturation value
with increasing time.  We should note that application of the bias
field to the system is sufficient to induce a large $Q$ value indicating
the existence of an ordered phase, and it also allows us to analogy
with static paramagnetic saturation  at large applied field values.
If one compares the our MC simulation findings with
the experimental data in Fig. 2 of Ref. \cite{Berger}, a well
agreement is seen between modeling and experiment.

\begin{figure}[!here]
\begin{center}
\includegraphics[width=12.0cm,height=7.5cm]{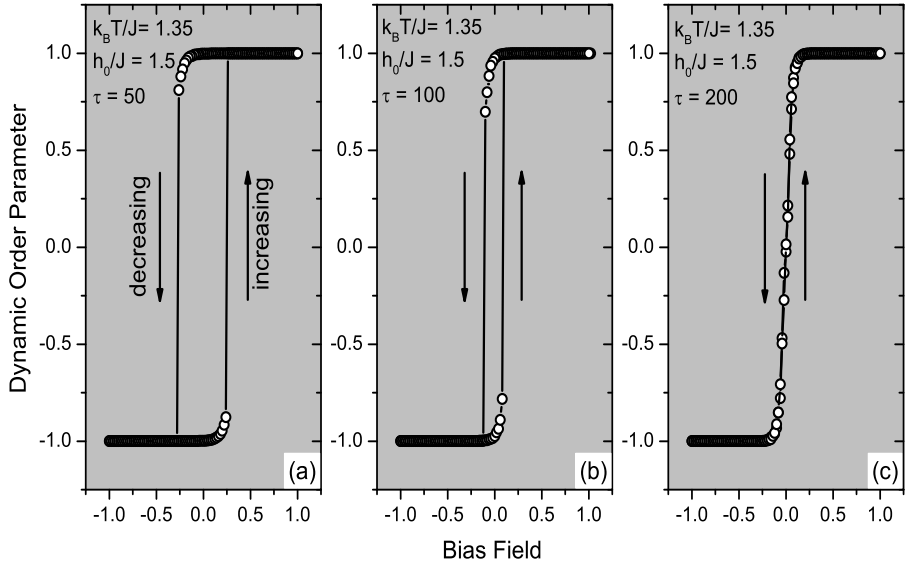}
\caption{Bias field dependencies of the dynamic order parameter  obtained for values
of $k_{B}T/J=1.35$ and $h_{0}/J=1.5$. The dynamic hysteresis curves are calculated for different values of
applied field periods: (a) $\tau=50$, (b) $\tau=100$ and (c) $\tau=200$.
The decreasing and increasing branches of the curves are denoted by downward and upward
arrows, respectively.}\label{Fig4}
\end{center}
\end{figure}

In the following analysis, we give the bias field dependencies of the dynamic order
parameter for a fixed temperature $k_{B}T/J=1.35$ and applied field
amplitude $h_{0}/J=1.5$ for varying applied field periods in Fig. \ref{Fig4}(a)-(c).
The measurement procedure to obtain the decreasing branches of the
hysteresis curves is as follows: The bias field starts at  $+1.0/J$, and then
it is slowly  decreased with a step  $\Delta h_{b}/J=-0.02$ until it reaches to $-1.0/J$.
We obtain the increasing branches of the curves using a similar way.
The decreasing and increasing branches of the hysteresis curves are
shown by downward and  upward arrows in figures.
One of the outstanding results is that dynamically ordered state is not
simply  inverted at $h_{b}/J=0.0$, and therefore,  $Q-h_{b}/J$ hysteresis
curve appears. When one compares the obtained curve with $M-H$ hysteresis curve for
ferromagnetic system at thermal equilibrium,  it may possible to say
that $h_{b}/J$ bias field appears to be the conjugate field of the
dynamic order parameter $Q$.  Furthermore, the aforementioned  behavior
strongly depends on the applied field period, and it disappears with increasing value
of the applied field period.

\begin{figure}[!here]
\begin{center}
\includegraphics[width=5.cm,height=6.cm,angle=270]{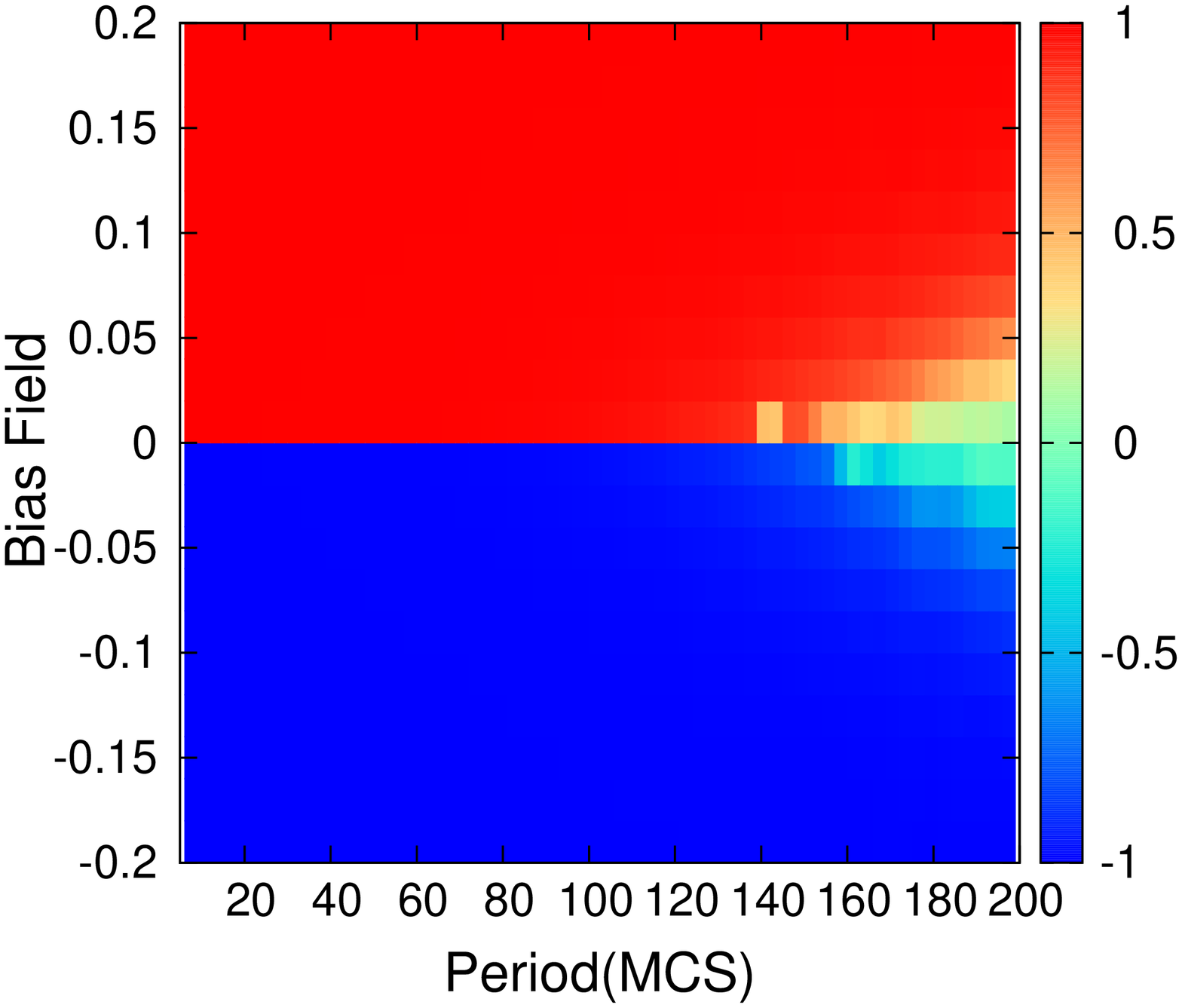}
\includegraphics[width=5.cm,height=6.cm,angle=270]{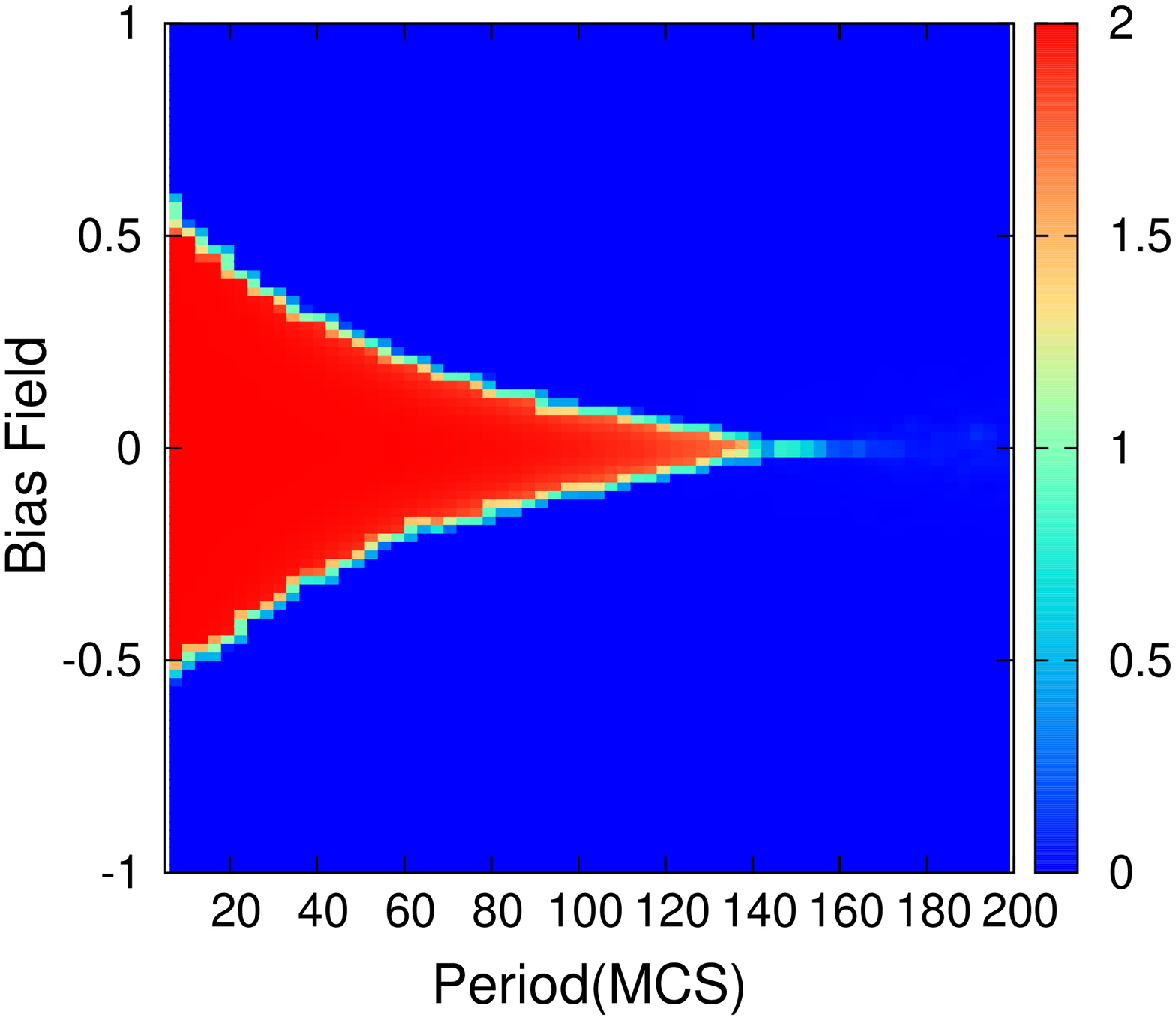}
\includegraphics[width=5.cm,height=6.cm,angle=270]{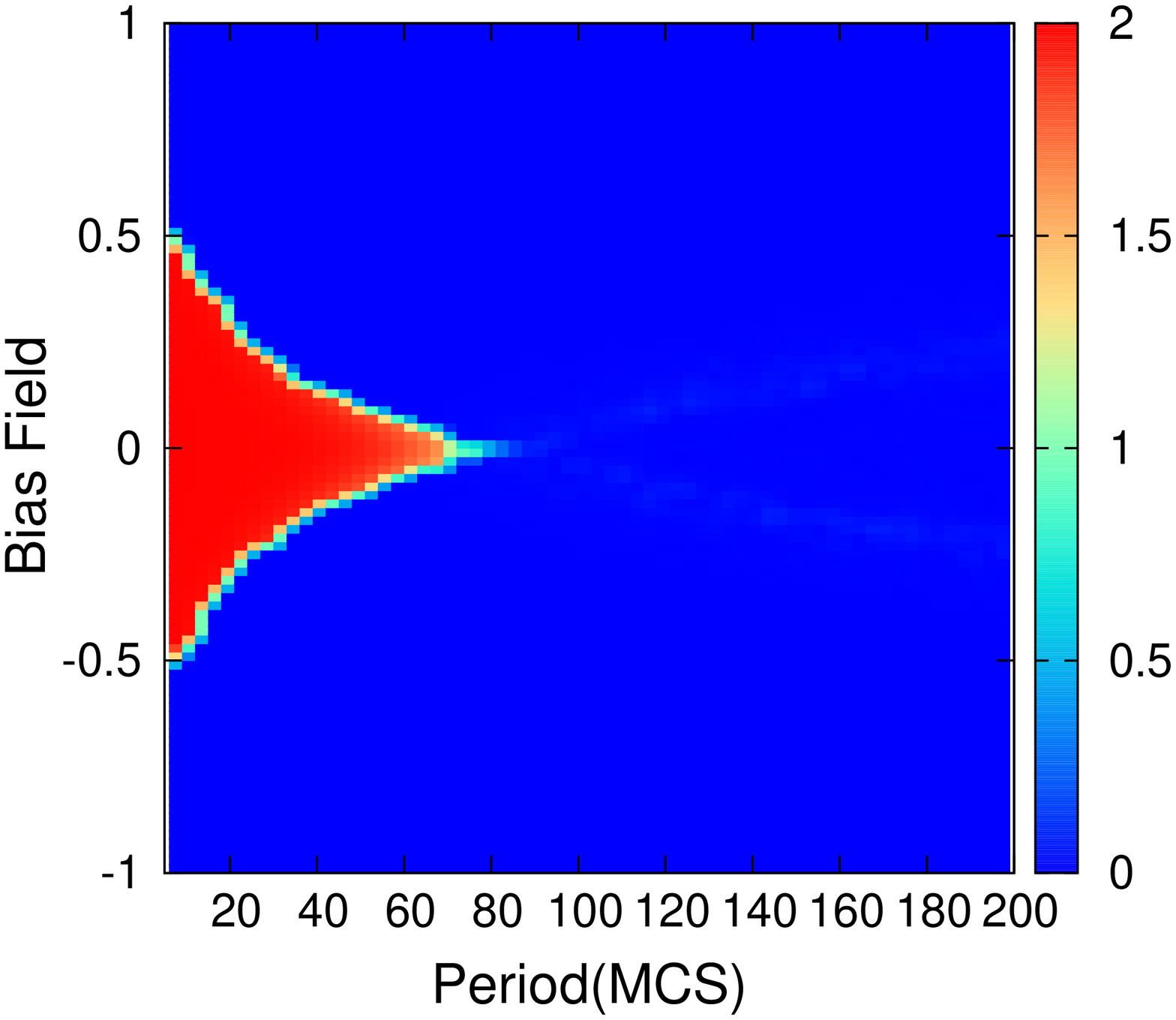}
\includegraphics[width=5.cm,height=6.cm,angle=270]{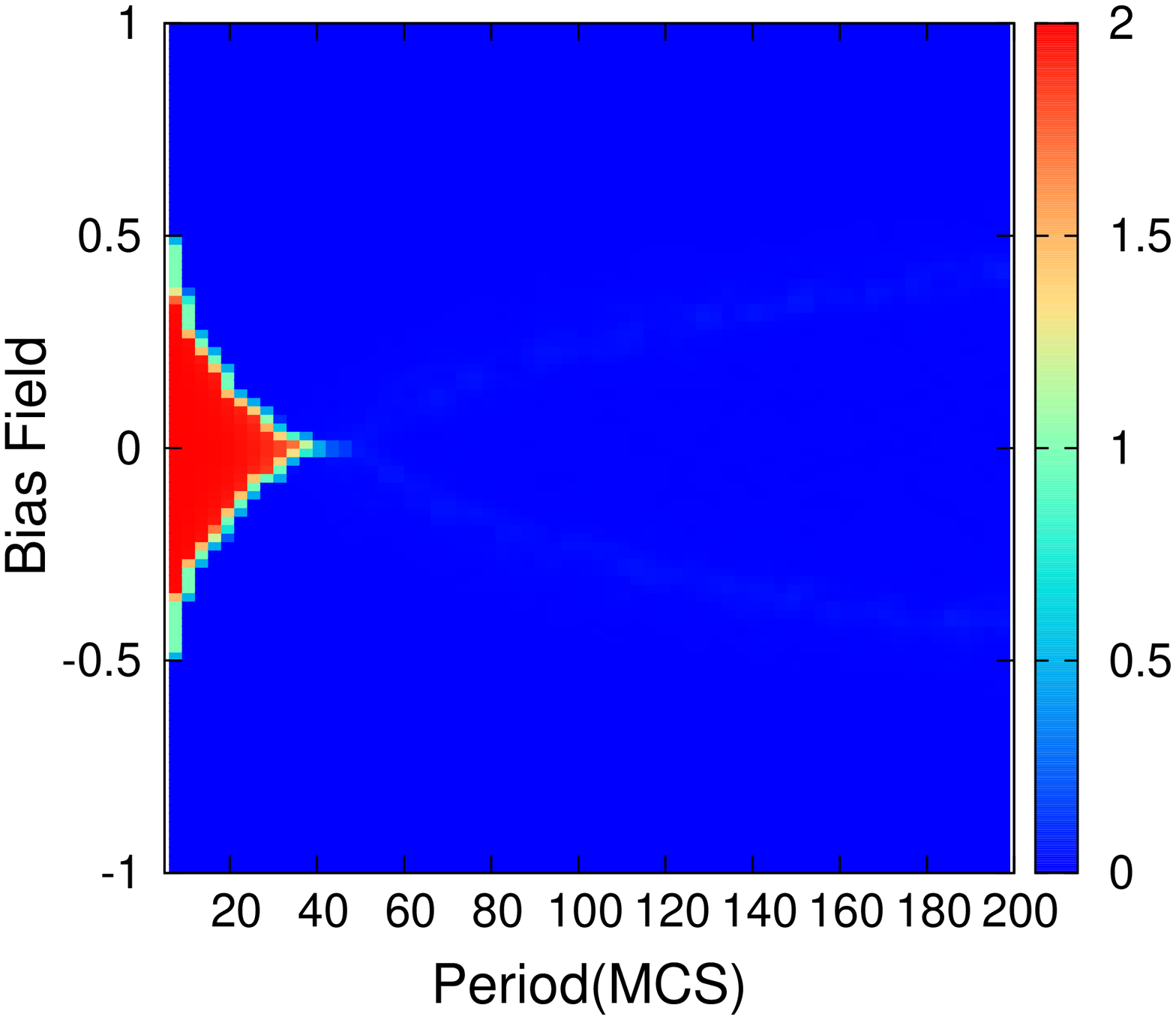}
\caption{For a fixed temperature value $k_{B}T/J=1.35$,
(a) displays the stable dynamic order
parameter for $h_{0}/J=1.5$ while (b-d) represent the variation of $\Delta Q$ in 2D contour representation
in a $(\tau-h_{b}/J)$ plane for varying applied field amplitudes such
as (b) $h_{0}/J=1.5$, (c) $h_{0}/J=1.7$ and  (d) $h_{0}/J=1.9$.}\label{Fig5}
\end{center}
\end{figure}

As a final investigation, we touch upon the period dependencies
of $h_{b}/J-Q$ dynamic hysteresis curve for a considered value
of temperature $k_{B}T/J=1.35$ with different values of the applied field amplitudes
in Fig. (\ref{Fig5}). In order to generate the Fig. \ref{Fig5}(a) constructed
as a color-coded map displaying the stable dynamic order parameter as a function
of the applied field period and bias field, we follow a similar way
used in Ref. \cite{Berger}. It is found that a sharp transition takes place
at $h_{b}/J=0.0$ between positive and negative $Q$ values up to a particular
value of $\tau$. With increasing value of $\tau$ starting from a relatively low value,
$Q$ vanishes at $h_{b}/J=0.0$. As a result of this, the sharp transition
behavior tends to disappear, and  there is no phase transition line as $\tau$ reaches
at a critical value. The curves in Figs. \ref{Fig5}(b)-(d) are plotted for three
values of applied field amplitude: (b) $h_{0}/J=1.5$, (c) $h_{0}/J=1.7$ and (d) $h_{0}/J=1.9$.
The figures represent a map  of $\Delta Q(h_{b}/J,\tau)=Q^{d}(h_{b}/J,\tau)- Q^{i}(h_{b}/J,\tau)$.
In other words, this formula determines  the differences of the $Q$ values obtained
for the decreasing branch $Q^{d}(h_{b}/J,\tau)$ and the increasing
branch $Q^{i}(h_{b}/J,\tau)$. It is clear from the figures that
the obtained $\Delta Q(h_{b}/J,\tau)$ curves appear as the roughly
triangular structure extending up to a critical period value  which
sensitively depends on the applied field amplitude and other system
parameters.  When the energy coming from the
time dependent applied field is increased, the boundaries of the
triangular shapes tend to show a decreasing trend, and it is possible
to mention that after a special value of the  amplitude of field, the triangular
shape completely disappears.

\section{Conclusion}\label{conclusion}
    In summary, we have investigated the thermal and magnetic phase transition features
of the ferromagnetic thin film system under the influence both bias field
and time-dependent magnetic fields. For this purpose, we have used Monte Carlo simulation
technique with single-spin flip Metropolis algorithm. First, to determine
the dynamic phase transition point, we have treated the thermal variation of
the heat capacity for considered values of the system parameters, and by benefiting from its peak,
we have constructed the dynamic phase diagram  in reduced magnetic field versus
temperature plane. It is reported that for a fixed value of $h_{0}/J$, as $\tau$ is increased,
the phase transition point is lowered  since decreasing field frequency gives rise to a decreasing
phase delay between the magnetization and magnetic field  and this makes the
occurrence of the dynamic phase transition  easy.  Next, we have focused on effects of the varying
bias field on the ferromagnetic thin film system, and we have studied the dynamic hysteresis behaviors as a
functions of system parameters in detail. It is emphasized that $h_{b}$ is
the conjugate field to the dynamic order parameter, and also $h_{b}$ appears to
define a phase line between two antiparallel dynamic ordered states.
As a final note, it should be underlined  that our MC
simulation results  obtained in the present work corroborate the experimental
observations found in Ref. \cite{Berger}.

\section{Acknowledgements}
The authors thank Andreas Berger from CIC Nanogune for critical reading
of the manuscript and for valuable  discussions on the subject. The numerical calculations
reported in this paper were performed at T\"{U}B\.{I}TAK ULAKB\.{I}M (Turkish agency), High Performance and
Grid Computing Center (TRUBA Resources).

\end{document}